\documentclass[aip,psfig,twocolumn,showpacs,superscriptaddress,reprint,floatfix]{revtex4-1}
\usepackage{amsmath}
\usepackage{amssymb}
\usepackage{color}
\usepackage{graphics}
\usepackage{epsfig}
\usepackage[normalem]{ulem}     
\usepackage{cancel}
\graphicspath{{Images_dusty_plasma/}}
\usepackage{xtab,afterpage,longtable}
\usepackage{booktabs}  
\usepackage[normalem]{ulem}
\usepackage{ltablex}      
\relax
\usepackage{float}
\begin{document}
\title{Shock wave bending around a dusty plasma void}
\author{Sachin Sharma}
\email{sachin.sharma@iitjammu.ac.in}
 \affiliation{Department of Physics, Indian Institute of Technology Jammu, Jammu, 181221, India}
 \author{Rauoof Wani}
 \affiliation{Department of Physics, Indian Institute of Technology Jammu, Jammu, 181221, India}
\author{Prabhakar Srivastav}
\affiliation{Institute for Plasma Research, Bhat, Gandhinagar, Gujarat, 382428, India}
\author{Meenakshee Sharma}
\affiliation{Independent researcher, Ahmedabad, India}
\thanks{Dr. M. Sharma contributed to experiments during her short visit to Shivalik Plasma Laboratory, IIT Jammu.}
\author{Sayak  Bose}
\affiliation{Columbia Astrophysics Laboratory, Columbia University, 550 West 120th Street, New York, NY 10027}
 \altaffiliation[Presently at ]{Princeton Plasma Physics Laboratory, Princeton, NJ 08540, USA.}

\author{Sanat Tiwari}
\affiliation{Department of Physics, Indian Institute of Technology Jammu, Jammu, 181221, India}
\email{sanat.tiwari@iitjammu.ac.in}
\author{Abhijit Sen}
\affiliation{Institute for Plasma Research, Bhat, Gandhinagar, Gujarat, 382428, India}
\date{\today}
\begin{abstract}
We report on experimental observations of the bending of a dust acoustic shock wave around a dust void region. This phenomenon occurs as a planar shock wavefront encounters a compressible obstacle in the form of a void whose size is larger than the wavelength of the wave. As they collide, the central portion of the wavefront, that is the first to touch the void, is blocked while the rest of the front continues to propagate, resulting in an inward bending of the shock wave. The bent shock wave eventually collapses, leading to the transient trapping of dust particles in the void. Subsequently, a Coulomb explosion of the trapped particles generates a bow shock. The experiments have been carried out in a DC glow discharge plasma, where the shock wave and the void are simultaneously created as self-excited modes of a three-dimensional dust cloud. The salient features of this phenomenon are reproduced in molecular dynamics simulations, which provide valuable insights into the underlying dynamics of this interaction.
\end{abstract}
\maketitle
\section{Introduction}
Compressional waves in dusty plasmas exhibit rich collective dynamics in laboratory experiments. These dust density waves, often referred to as dust acoustic waves (DAWs), arise due to a balance between dust inertia and plasma pressure~\citep{melzer2019physics}. Large amplitude waves can manifest as solitons, shocks, and rogue waves under the combined effects of nonlinearity, dispersion, and dissipation~\citep{Heinrich_2009_PRL,Garima_PRE_2021,tsai_lini_2016}. These waves can also exhibit nonlinear mixing, synchronization, and wave turbulence~\citep{Ruhunsiri_2012_PRE,Nosenko_2004_PRL,Zhdanov_2015_EPL}. When DAWs propagate through regions of density inhomogeneity, their phase velocity changes, a signature of refraction~\citep{Garima_2020_POP}. In the presence of an object with a size comparable to the wavelength, they exhibit diffraction phenomena~\citep{Kim_2008_POP}. Furthermore, DAWs are reflected when encountering a strong sheath or bias potentials~\citep{Krishan_2021_POP}. Collectively, DAWs provide a unique tool for visualizing various aspects of sound wave dynamics. In this paper, we report the first experimental observation of planar compressional wavefront blocking and bending around a dust void. In the present case, a self-excited nonlinear dust acoustic mode (a shock wave) interacts with a nonlinearly formed void. The experiment also reveals other novel features, such as the eventual collapse of the shock wave and the trapping of its constituent particles within the void, followed by a Coulomb explosion~\cite{Barkan_POP_1995b,Meyer_2016_IEEE,Ivlev_2013_PRE,Saxena_POP_2012,chaubey2022coulomb} that generates a bow shock~\cite{Meyer_2013_POP,Saitou_2012_PRL,Schutt_2025_POP}.\par

Dust acoustic shock (DAS) waves form because of a balance between strong nonlinearity and dissipation in a dusty plasma, in the presence of weak dispersion effects. DAS waves have been observed in many laboratory experiments in the past~\citep{Merlino_2012_POP,Heinrich_2009_PRL,Saitou_2012_PRL,Garima_2020_POP,Surabhi_2016_POP,Usachev_2014_NJP,Schutt_2025_POP}. Shocks assist in phase transitions \citep{Samsonov_2004_PRL}, analyzing the equation of state of strongly coupled plasmas \citep{Oxtoby_2013_PRL}, and display different particle patterns compared to the bulk~\citep{Kananovich_2025_PRE}. A simple way to generate a shock wave is to provide an impulse~\citep{Fortov_2005_PRE,Lin_2019_PRE}. It can be done mechanically using a piston or giving potential to a thin wire placed in the plasma~\citep{Kananovich_2020_PRE,Pustylink_2004_HT}.  In the present experiment, the shock is not generated by an externally imposed high-amplitude impulse but arises from self-excitation due to energetic ion streaming, giving rise to unstable DAWs~\citep{Heinrich_2009_PRL}. The excitation originates at one location, leading to the generation of compressional shock waves that then propagate in space towards the void structure and interact with it.

A void in the dust cloud is a dust-free region that emerges from the balance between the electrostatic and the ion drag forces ~\citep{Samsonov_1999_PRE, Morfill_1999_PRL, Goree_1999_PRE}. Such regions have been observed over an electrode (laboratory experiments under gravity) or at the cloud center (microgravity conditions) due to local ionization-induced instability in both microgravity and laboratory experiments~\citep{Morfill_1999_PRL,Samsonov_1999_PRE,Fedoseev_PRE_2015}. Past research has extensively focused on the formation of dust voids, their dependence on plasma parameters, and their steady-state equilibrium conditions~\citep{Yoshiko_POP_2018}. However, the interaction of a void with collective dust excitations such as DAWs or their nonlinear manifestations such as shocks or solitons has not been studied. What has been studied is the excitation of such structures when a flowing dust fluid encounters a solid obstacle immersed in the plasma ~\citep{Bailung_2020_POP,Meyer_2013_JPP,Morfill_2004_PRL}. Among the interesting results of these encounters are the diffraction of linear dust acoustic waves~\citep{Kim_2008_POP}, the formation of bow shocks \citep{Saitou_2012_PRL, Schutt_2025_POP,Meyer_2013_POP}, and the excitation of precursor solitons~\citep{Krishan_2024_POP}. In all such experiments, the obstacle consisted of rigid solid objects of various shapes and sizes. A self-excited dust void, as in the present experiment, is of a very different nature with a distinct dust density profile. It is composed of an almost dust-free central region that merges with the homogeneous dust density in the rest of the regions, with a steep density rise at the void boundary. Also, such voids remain soft, i.e., they shrink upon compression. Our present experiment, the first to the best of our knowledge, is devoted to studying the encounter of a dust acoustic shock wave with a soft compressible dust void. Both the shock wave and the void are self-excited as nonlinear structures of the system. \par

Our experiments were carried out in a DC discharge plasma with the dust cloud forming over the glass wall away from the electrodes in the diffused plasma region~\citep{Sachin_2023_AIPAdv}. The present study examines how a dust void modifies shock wave dynamics, leading to structural changes such as shock bending, void deformation, void-induced secondary wave structures, and Coulomb-explosion-like phenomena. We have further provided theoretical support to the experimental observations of the bending of shock waves by performing two-dimensional (2D) molecular dynamics simulations for parameters close to the experimental conditions. The simulations reproduce the salient features of the experiment. \par 

The manuscript is organized as follows. Section~\ref{setup} details the experimental dusty plasma setup, the plasma parameters calculated using the Langmuir probe,  and the dust parameters. Section~\ref{obs_an} discusses the experimental observations and analysis of the observed results. Section~\ref{md_sim} confirms our analysis of bending using the molecular dynamics simulation of the shock-void interaction, and Section~\ref{discuss} discusses various indirect tests to support the dynamical phenomenon. Finally, we conclude in sec~\ref{conc}.
\section{Experimental setup and dust cloud formation}
\label{setup}
\begin{figure}[t]
\includegraphics[width =\columnwidth]{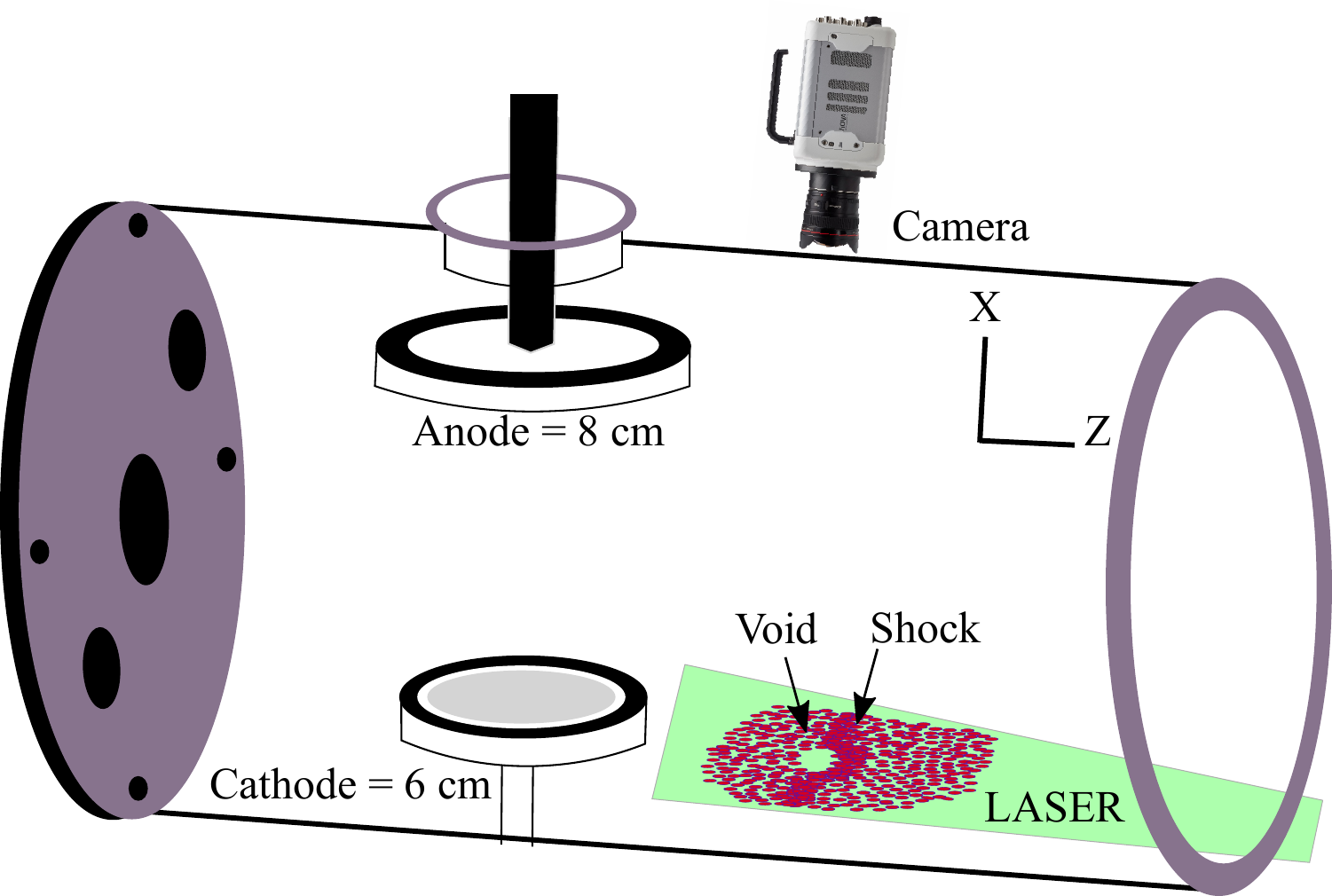}
\caption{Cartoon representation of the electrode assembly, the location of the dust cloud in the diffused plasma region, and the formation of a void within the dust cloud.}
\label{Fig_1}
\end{figure}

The experiments were performed in the Shivalik Plasma Device-I (SPD-I)~\citep{Sachin_2023_AIPAdv}. For complete details on the configuration, dimensions, and assembly of the device, readers are referred to ~\citet{Sachin_2023_AIPAdv}. A DC glow discharge plasma was created between two parallel plate electrodes facing each other. The grounded cathode and the anode were separated by $8$ cm, as shown in Fig.~\ref{Fig_1}. It is important to note that this asymmetric electrode assembly employs an anode that is larger than the cathode. This arrangement deviates from the conventional practice used in certain dusty plasma experiments~\citep{Hariprasad_POP_2018,DPEx2,Arora_2018_POP}, where a small anode and a large cathode are typically chosen to minimize ion streaming effects. We have found that for the conventional electrode configuration, the complex nonlinear interaction between shock waves and void formation does not take place, and it occurs only when a larger anode with a smaller cathode is used. We believe this configuration of electrodes enhances ion streaming, which in turn leads to the excitation of shock waves and the formation of a void in the dust cloud. A systematic investigation of various electrode configurations supporting this observation is presented in section~\ref{discuss}.
%
\begin{figure*}[!htb]
\includegraphics[width =\textwidth]{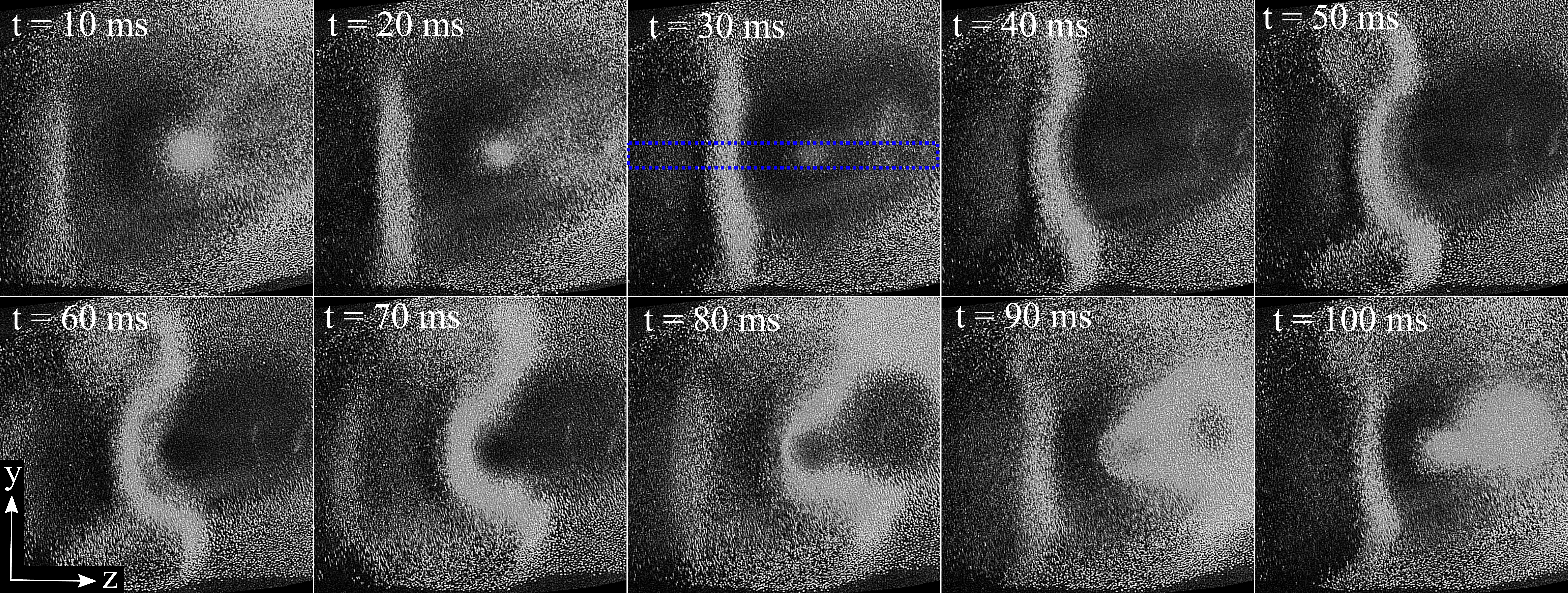}
\caption{The evolution of dust shock interaction with dust void. The central region of the shock wavefront bends as it propagates towards the void (10 to 80 ms) and then completely covers the void by transient particle trapping before creating a high-density dust region (90 - 100 ms).}
\label{Fig_2}
\end{figure*}


The experiments were performed in argon plasma at a working neutral pressure of 0.12 mbar (12 Pa), and the base pressure was $10^{-2}~\rm{mbar}$ (0.1 Pa). We introduced argon into the vacuum chamber through a port adjacent to the pumping port with a flow rate of 0.7 mg/s to prevent any directional flow of neutral gas into the experimental region of interest. This approach ensures that only diffused gas enters the system~\cite{dust_vortices_manjit_kaur, Manjit_PM_2016, Sachin_2023_AIPAdv}. To ensure high purity of the background gas before performing the experiments, we repeatedly purged the device with argon gas at high pressures $\left( \sim 2~\rm{mbar}~or~200 Pa\right)$ for at least 10 minutes before settling at the working pressure to conduct the experiment. Refer to ~\citet{Sachin_2023_AIPAdv} for more details on operation of the vacuum system. Plasma typically forms around a discharge voltage of $V_{\rm{D}} = 300$ V at this pressure.  However, experiments involving the formation of a void within the cloud (see the void in Fig.~\ref{Fig_1}) required higher discharge voltages of ($V_{\rm{D}} \approx 600$ V - $800$ V). \par

The location of the dusty plasma cloud is shown in Fig.~\ref{Fig_1}. The charged dust cloud forms in the diffused plasma region~\citep{Sachin_2023_AIPAdv,Minderhout_PSST_2020,Minderhout_JPD_2019,Abuyazid_2023_nature} over the glass surface of the device, far from the region between the parallel electrodes~\citep{Sachin_2023_AIPAdv,Sachin_2024_POP, Mangilal_RSI_2016}, as can be seen in Fig.~\ref{Fig_1}. In this region, the electron density can be two orders of magnitude lower than that between the electrodes~\cite{Sachin_2024_POP}. Hence, we refer to this region as the diffused plasma zone. The location, nature, and the physical properties of the cloud are highly reproducible and depend on the discharge voltage~\cite{Sachin_2023_AIPAdv}. The reason for this shifted location of the dust cloud is the expulsion of charged dust from above the cathode due to a primary void formation. The void imposes a strong outward force on these dust particles away from the cathode towards the glass walls. The falling dust particles eventually levitate in a region above the glass surface where gravity on dust particles is balanced by the upward electric field force due to the plasma sheath over the glass walls~\citep{Sachin_2023_AIPAdv}. The three-dimensional dust cloud was formed using monodisperse melamine formaldehyde (MF) particles of size 8 $\mu$m.  The lack of a 2D layer is an outcome of the formation of a potential well due to the geometry of the device. A detailed explanation has been provided by~\citet{Sachin_2023_AIPAdv}.\par

\section{Plasma and dusty plasma parameters}
The background plasma and dusty plasma parameters in the diffused plasma region are presented in Table~\ref{tab:parameters}. The electron density ($n_e$) and temperature ($T_e$) were measured using a single Langmuir probe biased by a high-voltage ramp~\cite{Bose_2017}. For MF particles with a mass density of $1.51$ gm/cc and dust size (d) of $7.93\pm0.12~\mu$m, the mass of the individual dust particles ($m_d$) was calculated to be $(3.92\pm0.18) \times 10^{-13}$ kg and the dust charge was estimated as $Q_d \approx -12766 \pm 724 ~e$ using the OML theory~\cite{melzer2019physics}, where $e$ is the elementary electron charge. \par

We also made an estimate of the typical dust density in the undisturbed region of the dust micro-particle cloud. Individual frames of the recorded video, captured using a PCO Edge 5.5 high-speed scientific CMOS camera (resolution: 2560 $\times$ 2160 pixels, frame rate: 100 fps), in the focused region were analyzed to obtain the dust density as $n_d \approx (7.1\pm1.4)\times 10^{10}$ m$^{-3}$. The dust-interparticle separation is found to be $a \approx 150\pm10~\mu$m.
\begin{table}[b]
\centering    \renewcommand{\arraystretch}{1.4} 
    \setlength{\tabcolsep}{5pt} 
    \begin{tabular}{lll}
        \hline
        \hline
        \textbf{Parameter} & \textbf{Experiment} & \textbf{Simulation} \\
        \hline
        $T_e$ & $2.3 \pm 0.1$ eV & --\\
        $n_e$ & (8$\pm$1) $\times$ 10$^{14}$ m$^{-3}$ & --\\
        $m_d$ & ($3.92\pm0.18)\times 10^{-13}$ kg & 4 $\times$ 10$^{-13}$ kg \\
        d & $7.93\pm0.12~\mu$m & Point particle \\
        $Q_d$ & $12766\pm724~e$ & $12766e$ \\
        $a$ & (150$\pm$10) $\mu$m & 140 $\mu$m \\
        $n_d$ & $(7.1\pm1.4)\times 10^{10}$ m$^{-3}$ & 1.62 $\times$ 10$^{7}$ m$^{-2}$\\
        $\lambda_d$ & 45$\pm$3 $\mu$m & 280 $\mu$m \\
        $T_d$ & 2.0$\pm$0.3 eV & 0.06 eV\\
        $\Gamma$ & 805$\pm$160 & 300 \\
        $\kappa$ & 3.3$\pm$0.3  &  0.5\\
        $\Gamma^*$ = $\Gamma e^{-\kappa}$ & 30$\pm$12 & 182 \\
        $\omega_{pd}$ & 291$\pm$33 Hz & 27.25 Hz \\
        $c_{d}$ & 1.3$\pm$0.2 cm/s & 0.76 cm/s \\
        $v_{sh}$ & 10-18 cm/s & 0.1 cm/s \\
        $a_v$ & 1-1.6 cm & 2 cm\\
        $\nu_{dn}$ & 8 s$^{-1}$ & --\\
        \hline
    \end{tabular} 
    \caption{Plasma and dusty plasma parameters for the experiment and the simulation.}
    \label{tab:parameters}
\end{table}
In addition, dust micro-particles are tracked using the open-source software ImageJ to obtain the position and velocities of dust particles~\cite{Schneider_2012_Imagej}. The root-mean-square velocity of the dust micro-particles is obtained by taking the full width at half-maxima of the velocity distribution to get a dust temperature of $T_d$ $\approx$ $2.0\pm0.3$ eV. These parameters allowed us to calculate the Coulomb coupling parameter $\Gamma$ = $805\pm160$. We also obtained the screening parameter $\kappa = a/\lambda_d = 3.3\pm0.3$, where the Debye length ($\lambda_d$) is estimated to be $45\pm3~\mu$m. The dust plasma frequency, is calculated to be $\omega_{pd}~\approx~291\pm33$ Hz. The value of the effective coupling parameter ($\Gamma^* = \Gamma e^{-\kappa}$) comes out to be 40
, which indicates that the dust component is in a strongly coupled liquid state \cite{Huang_2023_PRR}. The longitudinal dust sound speed, which is important in characterizing shocks, was found to be $c_d = 1.5$ cm/s. This value was obtained by using the expression $c_d = \lambda_d\omega_{pd}$. 
The damping rate due to dust-neutral collisions turns out to be 8 s$^{-1}$, calculated using the Epstein drag formula~\citep{Epstein_1924_PR}, given as
$
\nu_{dn} = \left( {\delta \, P}/{r_d \, \rho_d} \right) \sqrt{{2 m_n}/{\pi k_B T_n}}
$.
Here, 
 \( \nu_{dn} \) is the dust-neutral damping rate, \( \delta \) is the Epstein coefficient (typically \( \delta \approx 1.26 \) for argon), \( P \) is the neutral gas pressure, \( r_d \) is the dust particle radius,
\( \rho_d \) is the mass density of the dust particle, \( m_n \) is the mass of a neutral atom, \( k_B \) is the Boltzmann constant (\(1.38 \times 10^{-23} \, \text{J/K}\)), \( T_n \) is the neutral gas temperature. All calculated parameter values of plasma density, dust density, and temperature exhibit slight variability, which will also be reflected in the estimation of other parameters in the table~\ref{tab:parameters}. \par

The dust shock dynamics are observed at high discharge voltages around 700 V. The perturbation for the shock compression probably arises due to high ion streaming at such voltages. To investigate and validate the presence of plasma flow in the diffused regime, a directional probe was specifically designed and implemented. The probe consists of two planar, oppositely oriented electrodes, referred to as the upward-facing and downward-facing square electrodes, with identical surface dimensions of 5 mm × 5 mm. The electrodes of the probe are positioned vertically parallel to the discharge electrodes, with one facing the upstream direction of the flow and the other facing downstream relative to the presumed flow vector in the diffused plasma region.
The measured ion current densities from the two probe surfaces, denoted as $J_{\uparrow}$ and $J_{\downarrow}$, are used to compute the current ratio $J_{\downarrow}/J_{\uparrow}$. A ratio greater than unity ($J_{\downarrow}/J_{\uparrow}$ $>$ 1) is indicative of a net plasma flow directed downward, from the source region toward the lower part of the chamber, possibly terminating on the glass vessel walls. This directional inference is based on the standard Mach probe theory, where the asymmetry in the ion collection is directly related to the parallel component of the flow velocity~\citep{Chung_2012_mach}.
Furthermore, experimental observations reveal that the magnitude of the ion-current ratio increases with increasing discharge voltage. This trend suggests a corresponding enhancement in the plasma flow velocity, likely driven by stronger electric field gradients or increased ionization rates at higher applied voltages.
\begin{figure}[t]
\includegraphics[width =\columnwidth]{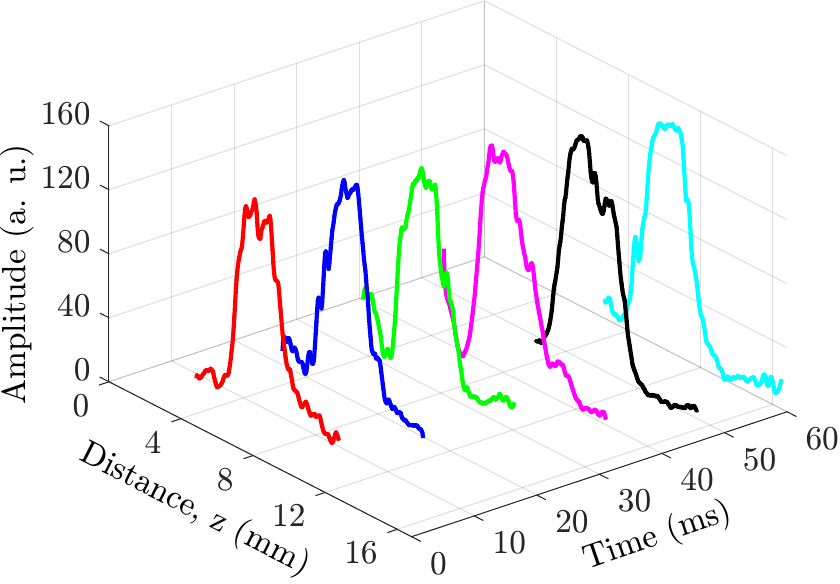}
\caption{Spatiotemporal evolution of the shock wavefront in the central region, obtained by averaging over the y-direction within the blue rectangular region shown in Fig.~\ref{Fig_2}, during its propagation toward the void.}
\label{Fig_3}
\end{figure}
\section{Experimental observations and analysis}
\label{obs_an}
The phenomenon of the interaction of the shock wave and the void has been observed around $V_D \approx 700$ V and $P=0.12$ mbar (12 Pa), when the anode-to-cathode size ratio is greater than unity. The phenomenon is highly reproducible, periodic in time, and is sustained as long as the plasma discharge conditions are maintained. Initially, the planar shock front advances uniformly towards the dust void region, causing compression of the dust particles in their propagation path.
\begin{figure}[b]
\includegraphics[width = \columnwidth]{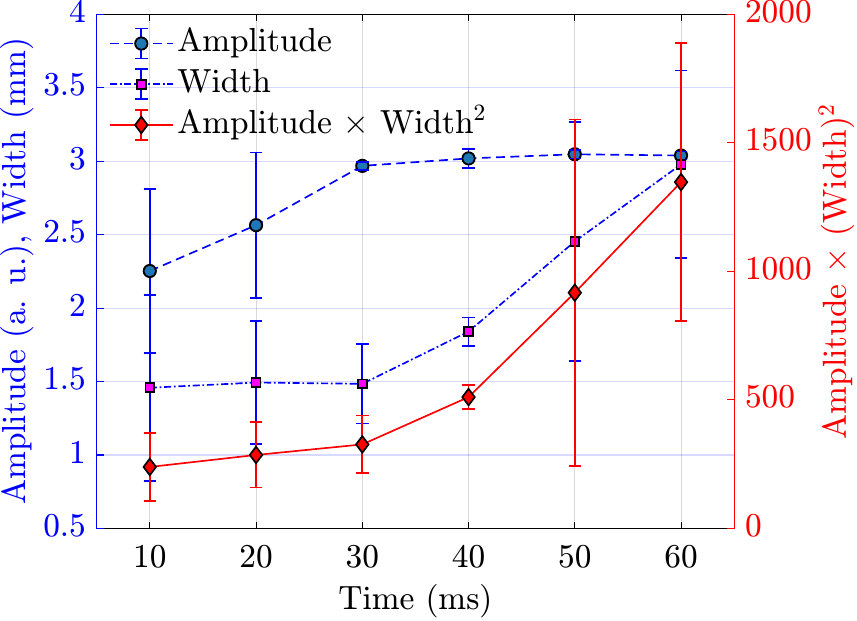}
\caption{Width and amplitude of the shock wavefront in the central rectangular region at different times as shown in Fig.~\ref{Fig_2}. The red markers represent the product (Amplitude $\times$ Width$^2$) at different times.}
\label{Fig_4}
\end{figure}
Figure~\ref{Fig_2} shows a wave compression moving from left to right in a series of images of the 2-D layer illuminated using a laser. At the initial time $t = 10$ ms, the wavefront remains planar and propagates towards the void. From $t = 30$ to $70$ ms, the wavefront starts to bend. This bending occurs when the central portion of the wavefront touches the dust void obstacle that resists the propagation through it, whereas the rest of the wavefront continues almost unperturbed. However, the dust void, a soft potential barrier, also shrinks during this process, leading to a slow skidding of the center part of the wavefront.
\begin{figure*}[t]
\includegraphics[width = \textwidth]{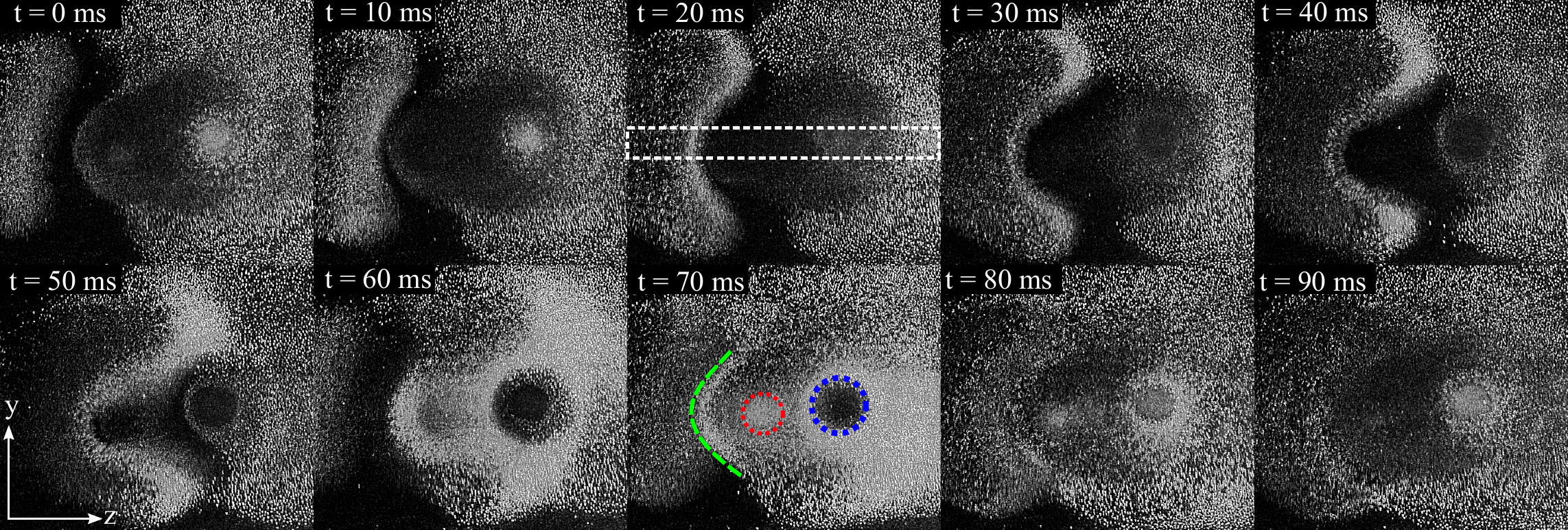}
\caption{Dust shock and void interaction in a 3D dust cloud in the bottom layer of the y-z plane. The top layer of the y-z plane, as shown in Fig.~\ref{Fig_2}, is $\delta$x  = 1 mm above this plane. Dust shock showing bending in the extended dust void region, leading to heap-void coexistence created by two polar regions: a heap (circled red) and a void (circled blue) through Coulomb explosion, also leading to bow-shock formation, shown by the curved line marked with green color.}
\label{Fig_5}
\end{figure*}
In general, even with a central curvature, the shock wave maintains its identity until $t = 70$ ms. Beyond $t = 70$ ms, as shown in Fig.~\ref{Fig_2}, the DAS structure collapses, leading to phenomena such as particle trapping and Coulomb explosion. Figure~\ref{Fig_3} shows the spatiotemporal evolution of the shock wavefront in a thin region at the center of the wavefront along the z-direction. We have observed that the wavefront maintains its amplitude, with a slight increase as it propagates for a duration of $60$ ms. This can also be followed in Fig.~\ref{Fig_4}. \par

The compressional mode plays an important role in the interaction with the void. To better understand whether it is a shock wave or a soliton, we carried out a few basic tests. Using the displacement of the highest-amplitude locations in time, we estimate the speed of the shock wave ($v_{sh}$) to be around $10-18$ cm/s  compared to the typical dust sound speed ($c_d$) of $1.5$ cm/s in the medium. It suggests that the nonlinear structure is moving with a Mach number greater than $6$. It is highly unlikely that a soliton moves with such a high Mach speed. However, we carried out the standard test of the amplitude $\times$ width$^2$ for these large-amplitude compressional waves. We found that the quantity amplitude $\times$ width$^2$ was not conserved during evolution as the product increased, as can be seen in Fig.~\ref{Fig_4}.

As the shock propagates, its width remains constant and the amplitude increases linearly till it reaches the dust void region (up to 30 ms as shown in Fig.~\ref{Fig_4}). As soon as the shock front reaches the dust void region, the shock width increases as a result of the sand piling effect at the void edge. Figure~\ref{Fig_4} also suggests that the shock amplitude remains constant after the shock front reaches the dust void region, in accordance with the earlier suggested sand piling effect.  \par

\begin{figure}[b]
\includegraphics[width =\columnwidth]{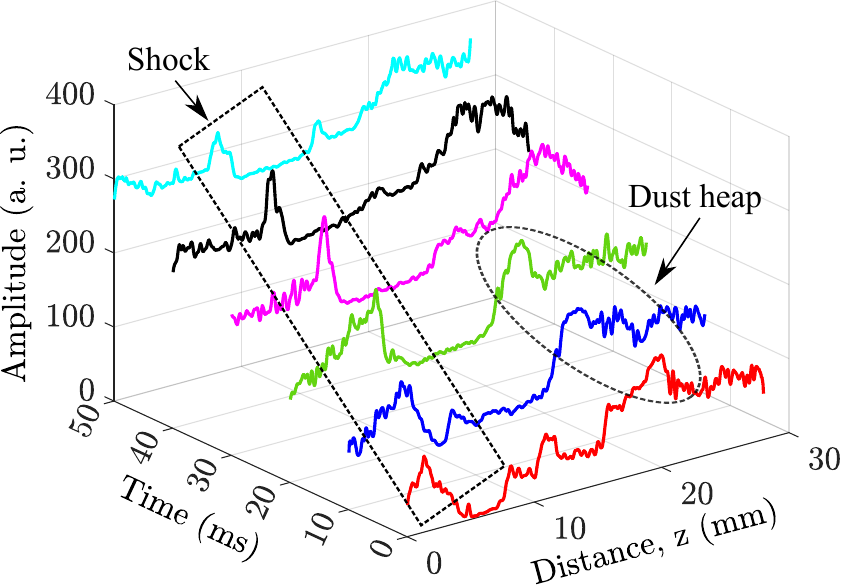}
\caption{Spatial intensity profiles of the shock wavefront along the z-direction (averaged over the y-direction of the white rectangular region shown in Fig.~\ref{Fig_5}) from t = 0 to 50 ms. Each curve is vertically shifted by an offset of 50 for clarity. The steep peak in the left region of the plot represents the shock wavefront, while the high-amplitude region on the right region corresponds to particle trapping and dust heaps.}
\label{Fig_6}
\end{figure}

So far, we have provided observations in a two-dimensional plane of an actual three-dimensional (3D) cloud. We further investigate the presence of similar interactions at multiple parallel layers. 
Figure~\ref{Fig_5} shows similar observations in the layer located one millimeter below the one shown in Fig.~\ref{Fig_2}. Specifically, the steep wavefront in the propagation direction and the gradual amplitude decay on the back side of the structure clearly reflect shock-like characteristics, as visible in Fig.~\ref{Fig_6}. The sharp peak with a steep fall, as shown in a rectangle in Fig.~\ref{Fig_6}, marks the location of the propagating shock front. On the right-hand side of the Fig.~\ref{Fig_6}, the elevated amplitude of the intensity profile corresponds to the dust heap structures as shown in an ellipse that develops due to the transient trapping of particles in the void. These features are identified by the characteristic sharp fall and subsequent rise in the intensity profile.\par 
\begin{figure}[t]
\includegraphics[width = 0.95\columnwidth]{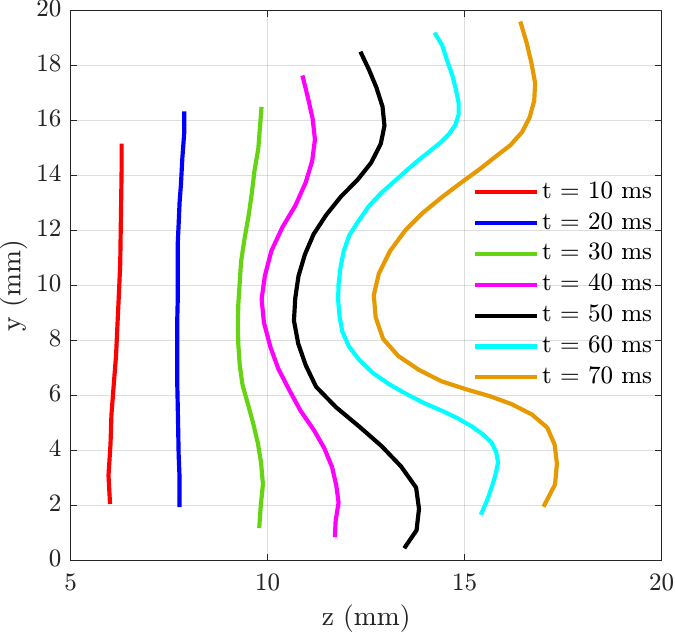}
\caption{2D spatiotemporal profile of dust shock wavefront drawn at different time frames showing bending with a decrease in shock speed in the extended dust void region.}
\label{Fig_7}
\end{figure}
\begin{figure*}[htb!]
\includegraphics[width =\textwidth]{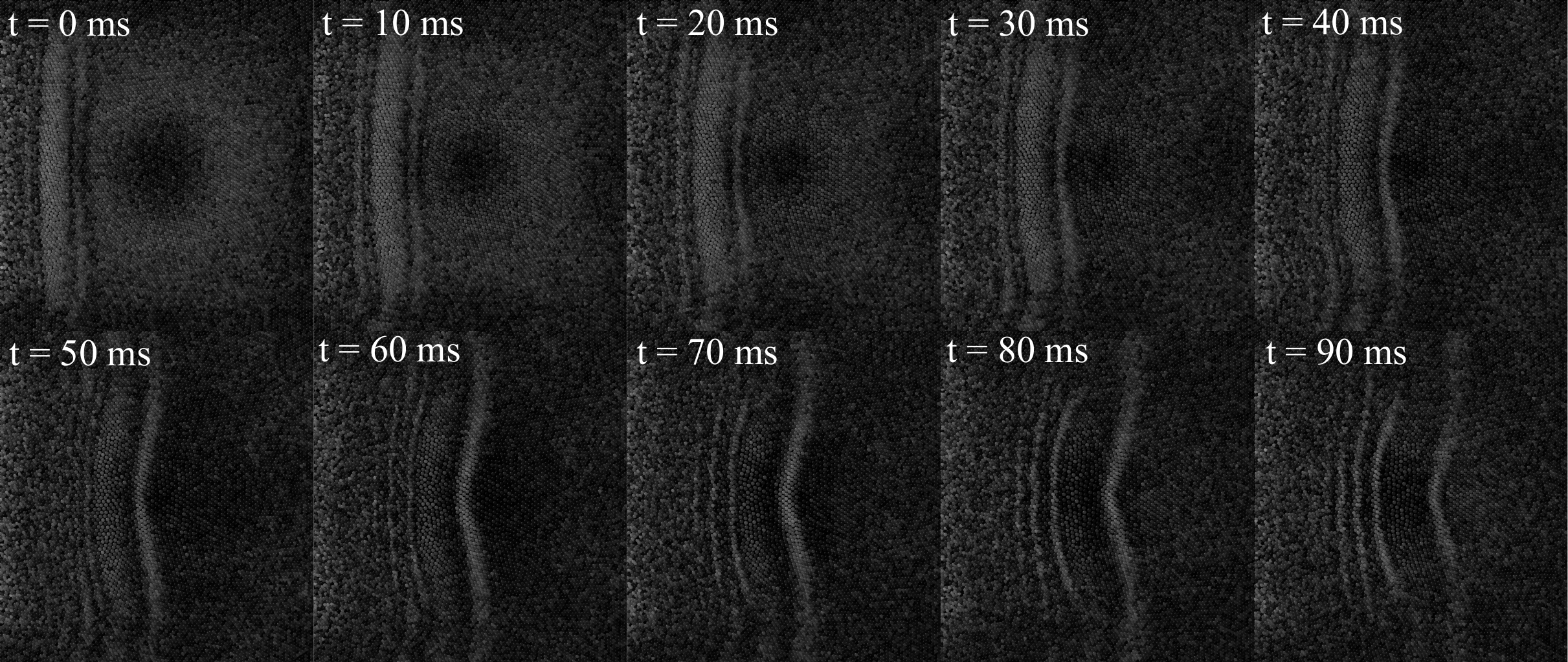}
\caption{Shock wave interaction with a dust void using two-dimensional molecular dynamics simulation.}
\label{Fig_8}
\end{figure*}

Figure~\ref{Fig_7} presents the smoothed density compression contours along the y-axis corresponding to the entire wavefront shown in Fig.~\ref{Fig_2}. Careful inspection suggests the following. 1) The shock wave front was initially planar and later bent from the center due to interaction with the soft potential barrier of the circular void region. 2) The soft potential barrier in the peripheral region of the dust void causes a slowing of the shock wavefront in the central region, rather than completely stopping the wavefront while the side fronts remain unaffected. 3) The central region of the wavefront keeps on propagating with a changed speed until it reaches an almost dust-free region (strong electrostatic field near the void center); after that, the wavefront collapses and loses its identity. This region is visible in the center of the void in Figs.~\ref{Fig_2} and \ref{Fig_5} at and around $t=50$ ms. \par

Some other interesting observations include the formation of a dust-void pair and dust heaps (heaps of dust particles). Dust heaps are regions of high local dust particle density in specific regions of the cloud. When the shock wave significantly bends around the void, it breaks, and the charged dust particles appear to get trapped transiently inside the void region, forming a dust heap in the middle of the void. These heaps are shown in Figs.~\ref{Fig_2} and \ref{Fig_5} in snapshots of 70 to 80 ms. \par

Although physical observations are the same in different layers (y-z plane) separated by a few millimeters, each layer has some distinct features. For example, for the middle and bottom layers, sometimes two clear voids coexist. As the shock wave propagates towards the right, the first void causes bending and eventually traps the particles, leaving the second void less affected. This leads to the emergence of a ``heap-void co-existence". \par

The transient trapping of charged dust particles in the void leads to a Coulomb explosion due to strong electrostatic repulsion~\cite{Barkan_POP_1995b,Saxena_POP_2012,chaubey2022coulomb}, as can be seen in Fig.~\ref{Fig_5} at $t=60 - 80$ ms. The explosion that occurred immediately after $t=60$ ms led to a bow shock formation, which is visually represented by the green curve at $t=70$ ms and quickly disperses outward. The symmetry of the bow shock is disturbed due to the presence of another void on one side of the heap, which opposes the outward propagation of the shock. \par

Although phenomena such as void-heap and void–void coexistence, Coulomb explosion, and bow shock are important nonlinear physical processes, their origin and evolution will be studied and presented separately in order to maintain the focus of this paper on the shock–void interaction.
\section{Molecular dynamics simulation}
\label{md_sim}
To support the above experimental findings, we performed molecular dynamics (MD) simulations using the open-source code LAMMPS ~\citep{LAMMPS_plimpton}. MD simulations have proved useful in the past to gain first-principles understanding of collective dynamics in dusty plasmas~\citep{Rauoof_RTT_PoP_2024,Dhaka_2024_POP}. Here we simulate identical point particles of mass \( m_d = 4 \times 10^{-13} \)~kg similar to those used in experiments. The interparticle interaction is taken to be the Debye-H\"{u}ckel or Yukawa potential:
\begin{equation}
\phi_{ij} = (q^2/4\pi \epsilon_0 r_{ij}) \mathrm{exp}(-r_{ij}/\lambda_d)
\end{equation}
The particle trajectories are obtained by solving the classical equations of motion $m \ddot{{r}}_i  =\ - \nabla \sum \phi_{ij}(r)$ where $r_{ij}$ is the distance between the $i^{th}$ and $j^{th}$ particles. We also point out that Langevin dynamics was intentionally avoided as its additional damping effect would have suppressed the shock front in the present system. \par

The dust particles are simulated inside a square box with $l_y=l_z$. The boundaries are periodic along the y-axis and absorbing along the z-direction. The simulation parameters are given in Table \ref{tab:parameters}. We start the simulation by randomly creating the dust particles inside the box at a particular temperature $T$, corresponding to the desired coupling parameter $\Gamma$. The system then evolves under equilibrium conditions for about $200~\omega_{pd}^{-1}$
using a Nosé-Hoover thermostat. During the equilibration of the system, the temperature and energy fluctuations are limited within $10^{-4} \%$ and $10^{-2} \%$, respectively.  \par

We then created a void in the form of an electric-field region at the center of the simulation box. The intensity of the electric field is such that it is maximum at the center of the void and decays with its radius. The radius of the void is $r_{\rm{c}}=0.4\times l_{\rm{y}}$, and the electric field varies as $E=E'\exp{([(y-y_{\rm{c}})^2+(z-z_{\rm{c}})^2]/r_{\rm{c}}^2)}$,
$E'=0.02E_0$ with $E_0=(k_{\rm{B}} T_{\rm{d}}/Q_{\rm{d}}\lambda_{\rm{d}})$. Here, $(y_{\rm{c}},z_{\rm{c}})$ is the center of the void, which is the same as the center of the simulation box. The electric field repels dust particles, creating a void. A compressional shock wave was excited along the z-direction at the left edge of the simulation box. The perturbation was applied to particles within a region spanning the entire length $l_\mathrm{y}$ in the $y$-direction and extending over $0.05 \times l_\mathrm{z}$ along the $z$-direction. The shock was initiated by imposing a one-time velocity impulse on the dust particles at the boundary, similar to the action of a piston, thereby generating a compressional front that subsequently steepened into a shock. \par

As the planar shock front approaches the void, it bends inward as the portion of the wavefront encountering the void edge experiences the obstacle and gradually slows down until it completely stops. The slowing of the wavefront is attributed to the softness of the formed void. This behavior is demonstrated in Fig.~\ref{Fig_8}. The results qualitatively reproduce the salient features of the bending of the shock wave observed in the experiment. In the simulations, the shock wave, despite bending, retains its shape because the electric field responsible for void formation is not very strong. Additionally, the simulation model of the void exhibits a gradual density depletion from the edge to the center, which differs from the experimental void that has a sharply defined edge. In addition, the simulation model does not include neutral drag effects. Though, the discrepancy arises in providing an exact experimental model in the simulation setup. However, the bending of the wave is demonstrated in alignment with experimental observations.
\section{Discussion}
\label{discuss}
Compressional waves traveling through a fluid are known to encounter processes such as diffraction, reflection, and refraction~\citep{Kim_2008_POP, Krishan_2021_POP, Garima_2020_POP}.
Usually, the bending of compressional waves is associated with diffraction phenomena, and such an observation has been reported to occur in dusty plasma if the wavelength of the compressional mode and the obstacle size are comparable~\citep{Kim_2008_POP}. In the present experiment, the dust shock wave propagates with a velocity of $100$ mm/s and a frequency of $11$ Hz, suggesting a typical wavelength of $10$ mm. The dust void edge that encounters the shock wavefront is about $a_v$ = $16$ mm in size and extends to almost twice this size in the wave propagation direction. Given that the obstacle size significantly exceeds the wavelength, the diffraction phenomenon is unlikely to be responsible for the observed wave bending in this study. \par

Another possible mechanism that causes wave bending is the phenomenon of refraction. We evaluated this as a possibility for the present experiment. We may note from Fig.~\ref{Fig_6} that the dusty plasma cloud density is a few times higher than that of the void. As suggested by \citet{Garima_2020_POP} in their experiments and as theoretically predicted by~\citet{Nishikawa_1975}, the phase velocity of the shock wave should have increased as it moves in the void region. However, our observations report otherwise and clearly demonstrate a reduction in the shock speed. \par

In general, the wave hitting large obstacles is likely to reflect. However, no such phenomenon is observed in the present experiment. This is due to the soft compressible nature of the void and the absence of rigid boundaries. As the wave reaches the periphery of the void, the void's compressible potential barrier de-accelerates the wave in the central region while the wave imparts momentum to the dust particles in the periphery region to counter the void's electric forces. The translation motion of the de-accelerated wave occurs until the wavefront reaches the deep void region, where the interplay of void forces completely overcomes the wave momentum and causes it to stop. \par

Another experimental feature that needs to be understood is physical mechanism responsible for the excitation of the compressional wave and the formation of the dust void. We believe that ion streaming is responsible for both the excitation and sustenance of the shock wave and the void. To confirm this hypothesis,  
we conducted an indirect test to see whether ion streaming-driven dust acoustic instability is the most probable cause. We carried out experiments with different pairs of electrodes as sources of plasma discharge. Table~\ref{tab:experiment} lists four such combinations. The cases with anode size larger than the cathode, i.e. $d_C < d_A$, are favorable for shock and void formations. We believe that the increased anode and smaller cathode sizes enhance ion streaming in the plasma medium.  \par
\begin{table}[h]
    \centering    \renewcommand{\arraystretch}{1.4} 
    \setlength{\tabcolsep}{8pt} 
    \begin{tabular}{c c c c}
        \hline
        Cathode & Anode & Condition & Dynamics \\
        $d_C$ (cm) & $d_A$ (cm) &~& ~ \\
        \hline
        8  & 8 & $d_C > d_A$ & No shock/void \\
        8  & 6 & $d_C > d_A$ & No shock/void  \\
        8 & 5 & $d_C > d_A$ & No shock/void \\
        6 & 8 & $d_C < d_A$ & Shock and void \\
        5 & 8 & $d_C < d_A$ & Shock, void \\
        \hline
    \end{tabular} 
    \caption{Conditions of electrodes configuration for Shock-void observation}
    \label{tab:experiment}
\end{table}
All these dynamics were observed in a diffused plasma region characterized by negligible visible glow, making direct optical measurement of local glow intensity fluctuations challenging. However, we made Mach probe measurements in the diffused plasma region to support the increased ion streaming statements. Measurements indicate an increased ion current towards the glass surface. In addition, the ion current increases with increasing discharge voltage. It clearly suggests that the ion streaming is directed towards the glass surface. Although the exact ion streaming velocity could not be determined due to calibration limitations of the Mach probe, the measured ion current values strongly support the observed behavior. \par

The observed dust acoustic shock wave propagates in a low-pressure regime where the dust-neutral collision frequency is estimated to be $\sim$ 8 Hz, while the wave frequency is $f  = 11$ Hz. This condition $\nu_d < f$ supports under damped wave propagation, allowing the shock structure to form and evolve before dissipative effects dominate. Also, the whole interaction dynamics takes place up to a length scale of around 10 mm while the damping length is calculated to be $l_d$ = 12.5 mm. This indicates that the wave amplitude (and energy) will not decay due to neutral drag throughout its evolution, explaining the dissipation-less nature of the shock wave. In addition, the explanation suggested by Fortov et al. and Ivlev et al. fits reasonably well~\citep{ivlev2010waves,Fortov_2004_PRE}. The instability-assisted energy flux appears to sustain and initially increase the compressional shock amplitude in our experiment against dissipation. Possible mechanisms include dust acoustic and modulational instabilities. \par
\section{Summary and Future-scope}
\label{conc}
The present work investigates the interaction of a self-excited dust acoustic shock with a soft dust void in a dusty plasma cloud. In the experiment, at high discharge voltages, a dusty plasma cloud formed in the diffused plasma region exhibits complex, self-excited, and steady state nonlinear dynamics. These dynamics include the periodic interaction of two nonlinear structures, a shock, and a void. In the later stages of interaction with the void, the shock bends significantly and loses its identity as a wave, releasing particles from its collective dynamics. These particles transiently get trapped within the void before the accumulated particle cloud undergoes a Coulomb explosion, forming a bow shock. The observed bending phenomenon of the shock wave is attributed to the relative size of the void compared to the wavelength of the wave and the soft nature of the void as an obstacle. These observations present a detailed visible record of the full life cycle of a DAS wave from its build-up as a collective oscillation to its collapse. \par

While the present study primarily presents a two-dimensional view of the dust cloud, the actual interaction between the DAS wave and the void is inherently three-dimensional. Although the interaction is typically symmetric along the x-axis, the complete three-dimensional dynamics involves additional effects such as wave reflection and the formation of multiple voids along the depth (x-direction). To explore these aspects, a tomographic reconstruction is being prepared using multiple y–z, and x-z layer images. Further, the force or effective potential in the dust void and heap regions will also be explored. These studies, to be reported in the future, will complement the present findings by providing a more complete picture of the interaction dynamics in the dust cloud.
\section*{Supplementary Material}
The supplementary material consists of three sections: Section I includes the schematic of the directional probe and data values. Section II includes a better visualization of the dust heap-void and bow shock formation. Section III includes the video links to shock-void interaction.
\section*{Acknowledgments}
 ST acknowledges support from the IIT Jammu PRAISE grant PRA-100003 for the present work. AS is grateful to the Indian National Science Academy (INSA) for the INSA Honorary Scientist position.
\bibliography{dustyplasma}
\end{document}